# Preparation and characterization of Bismaleimide resin/titania nanocomposites via sol-gel process

Guotao Lu[a,b,*] Ying Huang[b]

[a]*Research and Development Center, Bonna-Agela Technologies, Inc. Wilmington, DE 19808, USA*
[b]*Center for Molecular Sciences, Institute of Chemistry, Chinese Academy of Sciences, Beijing 100080, People's Republic of China*

**Abstract:** Bismaleimide (BMI) resin/ titania nanocomposites were synthesized from allylated-phenolic modified bismaleimide resin and $TiO_2$ via the sol-gel process of tetrabutyltitanate ($Ti(O^nBu)_4$, TBT). These nanocomposite materials were characterized by FT-IR, XRD, FE-SEM, TGA and DMA. It was found that the nano-scale $TiO_2$ particles were formed in the AP-BMI resin matrix, and the average primary particle size of the dispersed phase in the nanocomposites was less than 100nm, but the particle aggregates with bigger size existed. Obvious improvements of glass transition temperature and heat resistance properties of the AP-BMI resins were achieved by the introduction of nano-sized $TiO_2$ inorganic phase, and the modulus of the material was also improved.

## 1. Introduction

There has been increasing interest in Organic-Inorganic Nanocomposite (OINC) materials due to their wide and potential applications in electronics, optics, chemistry, and biomedicine [1-4]. The incorporation of inorganic component into organic polymer matrix has greatly improved the thermal and mechanical properties of the polymer. And careful design of the inorganic phase can introduce many new properties into the polymers, such as catalysis [5], photo-refraction [6], electroluminescence [7], and photo-chromophore [8], etc. The commonly used inorganic phases are nano-scale metal particles [9], metal oxides [10], semiconductor nanocrystals [11], carbon nanotubes [12], and mesoporous organosilica [1].The sol-gel process, which produces glasses and ceramics at a relatively low temperature rather than at very high temperature needed for conventional inorganic glasses and ceramics, is a convenient method to be used to prepare OINC [10]. Soluble polymers or their monomers are usually introduced into the sol-gel system to provide the organic component in OINC [10]. Organic-inorganic nanocomposite materials of polymethacrylates [13], polyacrylonitrile [14], polystyrene [15], polycarbonate [16], epoxy resin [17], phenolic resin [18] and polymaleimide [19] with $SiO_2$ or $TiO_2$ nano-particles have been prepared through the sol-gel process.

The resultant nanocomposites can be divided into two classes according to the interfacial nature: one is nanocomposites with interfacial covalent bonds or ionic-covalent bonds between the organic and inorganic phases [13, 15]; the other is nanocomposites without covalent bonds between the two phases, only with hydrogen bonds or Van Der Walls interaction between them [20-22]. Generally, the nanocomposite material with covalent bonds between the two phases has better thermal and mechanical properties than those without covalent bonds [23].

High performance thermosetting resins have wide applications in many fields such as aerospace and semiconductor industries. To improve their properties further, nanocomposite materials based on high performance thermosetting resins have been studied, such as PMR-15 with clay [24], polybenzoxazine with clay [25] and $TiO_2$ [26], Novolac resin with $SiO_2$ [27] and $TiO_2$ [28,47], and cyanate ester with clay[29], etc.

Bismaleimide resin is one of the most important classes of high performance thermosetting resins, with its outstanding properties including high tensile strength and modulus, excellent heat, chemical and corrosion resistance [30, 31]. However, to overcome its brittleness (the major disadvantage) and improve its thermo-mechanical properties, the performance enhancement is needed.

Recently, various works have been done to overcome BMI resins' shortcoming and enhance its thermal and mechanical properties, for example, the studies of BMI-containing novel monomers [32-33], its copolymers [34-35], polymer alloys [36-39], carbon fiber [40-41] and Al-B whiskers [42] reinforced composites.

Performance enhancement of BMI resin has also been studied by preparing its nanocomposite material with clay [43] and $TiO_2$ (polyetherimide as BMI's modifier) [44]. In the present work we reported the preparation of BMI resin/$TiO_2$ nanocomposites by the sol-gel process and studied the morphology, thermal and mechanical properties of the resultant materials. To improve the toughness and reduce the viscosity of BMI melt, allylated novolac phenolic resin (AP) has been used as BMI's modifier.

## 2. Experimental

### 2.1. Materials.

Phenol (CP) and formaldehyde (37% in water, CP) were from Beijing Organic Chemicals Factory (China), and were used as supplied. Allyl chloride was obtained from Qilu Petrochemical Co. (China) and used after distillation. 4,4′-Bismaleimidodiphenyl methane (BMI) was purchased from Fenguang Chemical Co., Ltd. (China). It is a crystalline substance with melting point of 151-154°C and the purity greater than 99%. Tetrabutyl titanate ($Ti(O^nBu)_4$, TBT) was a chemical pure reagent from Tianjian 2nd Chemicals Factory (China), and was used as supplied. Acetyl acetone (ACAC), Tetrahydrofuran (THF) and n-Butanol (BuOH) were analytical reagents from the Beijing Chemical Factory (China). THF was purified by reflux over $Na/CO(C_6H_5)_2$ and distillation. De-ionized water was acidified by hydrochloric acid (35-38%) to pH=2.

The allylated novolak resin (AP) was prepared according to the method described in the reference [45]. Molecular weight of AP was 1075, and the degree of allylation was 59% determined by $^1$H-NMR.

### 2.2. Synthesis of The ally-modified novolak resin-BMI prepolymers (AP-BMI).

The allylated Phenol resins (AP) were prepared according to the literature [45]. To synthesize AP-BMI, 61.0g of APN (containing 0.25mol allyl groups) was charged in a three-necked flask and heated to 120°C. Then, 44.8g BMI (0.25mol) was introduced. The reaction was continued for 2hrs to give a transparent red-brown AP-BMI prepolymer. The formation of AP-BMI was also confirmed by FT-IR and $^1$H-NMR ($CD_3COCD_3$) spectra, and will be discussed in discussion part.

*Email: luguotao@gmail.com.

## 2.3. Synthesis of the AP-BMI-2/TiO$_2$ nanocomposites.

AP-BMI/TiO$_2$ nanocomposite materials were prepared by using the following procedure: 8.16g BMI-AP (Allylation degree is 59.2%) was weighed into a three-necked round flask equipped with a mechanical stirrer and a reflux condenser. Then 6.23g THF was added into the flask and the system was heated in an oil bath to keep the mixtures' temperature at 80°C. After the prepolymer was dissolved completely, the clear mixture of 0.68g Ti(O$^n$Bu)$_4$ and 0.3g acetyl acetone was added to the solution, and at last 0.03g acidified water with pH=2.0 was added to the mixture to perform the sol-gel reaction. The reaction time was 1.5 hours and the system's temperature was kept at 80°C. Then the volatile was removed at 50-130°C under reduced pressure to give the mold powder of the nanocomposite.

Specimens of the nanocomposites were prepared by compression molding. The curing cycle was 140°C/20MPa/4h + 200°C/20MPa/6h. The post cure schedule was 250°C/6h. The compositions of the AP-BMI/TiO$_2$ nanocomposites are listed in Table 1.

**Table 1.** Compositions of the AP-BMI resin/TiO$_2$ nanocomposites.

| Code | TiO$_2$ content (wt-%) | AP (wt-%) | BMI (wt-%) | AP/BMI by wt-% |
|---|---|---|---|---|
| AP-BMI | 0 | 54.6 | 45.4 | 0.6:1.0 |
| AP-BMI/Tit-2 | 2 | 53.5 | 43.5 | 0.6:1.0 |
| AP-BMI/Tit-5 | 5 | 51.9 | 43.1 | 0.6:1:0 |

## 2.4. Measurements.

$^1$H NMR spectra were obtained with a Bruker MW 300 spectrometer with CDCl$_3$ as solvent. FT-IR spectra were recorded on a Perkin-Elmer model 1600 IR spectrometer. X-ray diffraction diffraction measurements was examined with a high-resolution Philips X'Pert MRD diffractometer. For the Bragg-Brentano scan, the primary optics module was a combination Gobel mirror and a two-crystal Ge(220) two-bounce hybrid monochromator which produces pure CuK$\alpha_1$ radiation ($\lambda$ = 0.1540562 nm) with a divergence of 25 arc sec. A 0.18º parallel plate collimator served as the secondary optics. Dynamical mechanical analysis (DMA) was performed on a Perkin-Elmer DMA-7 Dynamical Mechanical Analyser in the bending mode with the specimen dimension 15mm×4mm×1.5mm. The measurements were conducted at 2Hz in the temperature range from 40°C to 350°C at heating rate 20°C/min. Thermogravimetric analysis (TGA) was performed on a Perkin-Elmer TGA-7 thermogravimeter with a heating rate of 20°C/min under a nitrogen flow rate of 100 ml/min. The microstructure features of the nanocomposite materials were examined with a Hitachi S-900 field emission scanning electron microscope. The fracture surfaces of the samples were coated with gold to eliminate charging effects, and a high voltage (10KV) was used.

## 3. Results and discussion

### 3.1. Synthesis of the AP-BMI/TiO$_2$ nanocomposites.

The synthesis of AP-BMI prepolymer resin was monitored by FT-IR and $^1$H-NMR spectra. Fig.1 shows $^1$H-NMR (CDCl$_3$, 300M) of AP-BMI prepopymers. The structure of the AP-BMI prepolymer is shown as following:

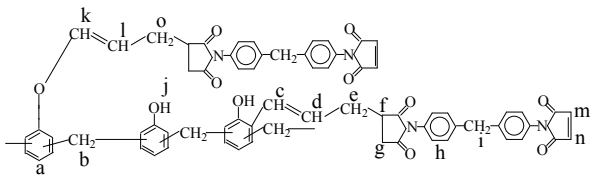

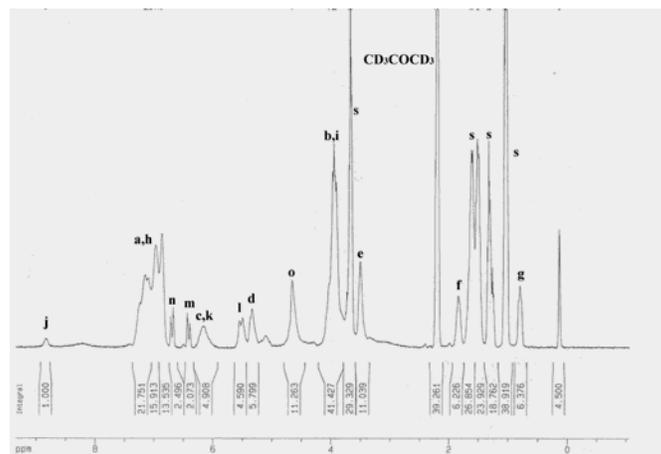

Figure 1. $^1$H-NMR (CD$_3$COCD$_3$, 300M) of AP-BMI prepopymers (in the figure, "**s**" reprents peaks from solvent CH$_3$CH$_2$CH$_2$CH$_2$OH).

Table 2. Chemical shifts of different H in AP-BMI prepolymer

| Symbol | a | b | c | d | e | f | g | h |
|---|---|---|---|---|---|---|---|---|
| Chemical shift (ppm) | 6.5-7.3 | 3.5-4.0 | 6.20 | 4.95-5.15 | 3.2-3.4 | 1.8 | 0.8 | 6.5-7.3 |

| Symbol | i | j | k | l | m | n | o |
|---|---|---|---|---|---|---|---|
| Chemical shift (ppm) | 3.5-4.0 | 8.95 | 6.20 | 5.3-5.5 | 6.4 | 6.7 | 4.3-4.8 |

(Unit: ppm)

The chemical shifts of all the different type Hydrogen were summarized in Table 2. The molecular structure of AP-BMI prepolymer was confirmed by the presence of all the characteristic peaks and peak integration. More proof is afforded by FT-IR spectroscopy of AP-BMI prepolymer, which is shown in Fig. 2 (a), and It can be seen from the figure that appearance of the carbonyl bands in BMI ring at 1770 and 1711cm$^{-1}$, and the peaks from the end double bonds R–CH=CH$_2$ at 990cm$^{-1}$ and 930cm$^{-1}$ have decreased a lot compared with FT-IR spectroscopy of AP, furthermore, the peak at 930cm$^{-1}$ have been split into two peaks at 920cm$^{-1}$ and 950cm$^{-1}$ which shows the presence of two kinds of double bonds, one is at the chain end R-CH=CH$_2$, the other is in the middle of chain R-CH=CH-R'. Those results further confirmed the structure of AP-BMI prepolymer shown above.

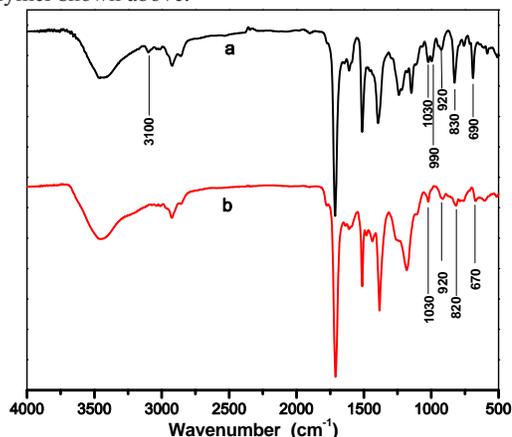

Figure 2. FT-IR spectra for the AP-BMI resin/TiO$_2$ nanocomposites: (a) AP-BMI; and (b) AP-BMI/Tit-2.

The synthesis process of the AP-BMI/TiO$_2$ nanocomposites has been shown in Scheme 1. To reduce the hydrolysis rate of tetrabutyltitanate (TBT), AcAc has been used to complex with Ti atoms. The hydrolysis of TBT gives Ti-OH groups, which condensed into TiO$_2$ nano-particles at BMI resin matrix. From Scheme 1, we can see that just part of TiO$_2$ nano-particles are covalent bonded into BMI resin network, and others are not bonded into the resin network, just dispersed in the resin matrix.

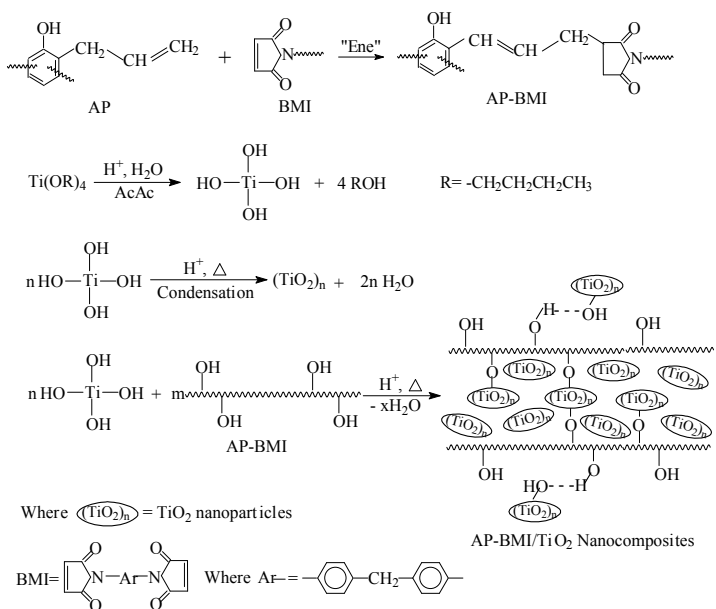

Scheme 1. The preparation of AP-BMI/TiO$_2$ nanocomposites.

Figure 2 and 3 give the FT-IR spectra for the AP-BMI prepolymer (a) and the corresponding nanocomposite AP-BMI/TiO$_2$ (b). The formation of TiO$_2$ particles can be proved by the appearance of a new adsorption band at 670 cm$^{-1}$ (Fig. 3c) from Ti-O-Ti in TiO$_2$, whereas the presence of Ti-O-C peak at 827 cm$^{-1}$ (Fig. 3c) shows that the hydrolysis of TBT is not complete. The formation of TiO$_2$ nanoparticles can also be seen from XRD and SEM study.

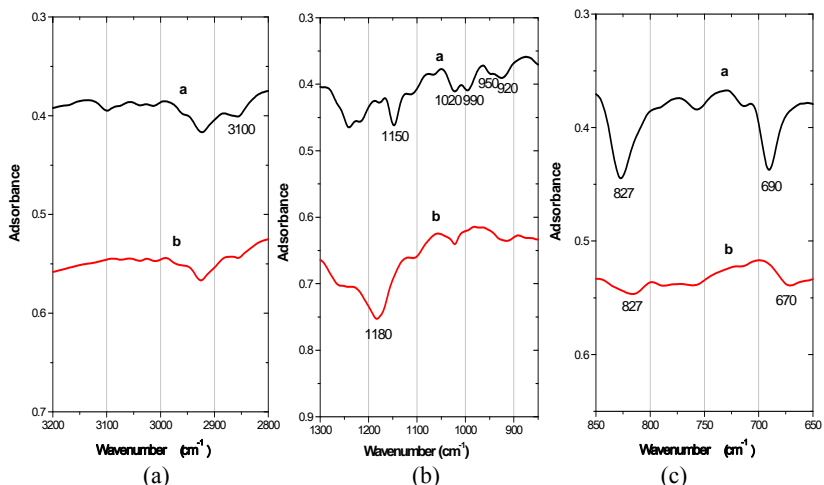

Figure 3. Magnification of Figure 2.

The proposed curing mechanism of the AP-BMI/TiO$_2$ nanocomposites is similar to DABA-BMI system given by Sung and Phelan (Figure 1 and 2 there), and it includes Ene, Diels-Alder, homopolymerization, re-Aromatization, and alternating copolymerization [46]. Those reactions can be proceeded by the disappearance of the absorptions at 3100cm$^{-1}$(Fig. 3a), the significant reduction of the bands from allyl groups at 990 and 930cm$^{-1}$(Fig. 3c) and the disappearance of the absorption at 950cm$^{-1}$ (Fig. 3c) from the double bonds in BMI ring, which also resulted in the shift of the carbonyl band from 1711 to 1706cm$^{-1}$ (Fig. 3c). Based on above results and the fact that the BMI resin was cured at 250°C, we believe that the AP-BMI resin may have been cured completely, according to the remarks by Phelan and Sung [46]. This accounts for the good thermal and mechanical properties of the composite resin.

Fig. 4 shows XRD pattern of the AP-BMI/Tit-5 nanocomposites. The XRD pattern can be identified as Anatase TiO$_2$. The peak broadening for every peak is due to the crystalline size of TiO$_2$ in nanometer size. This is consistent with conclusion from FT-IR studies with the formation of TiO$_2$, and SEM results which show the presence of TiO$_2$ particles of less than 100nm size.

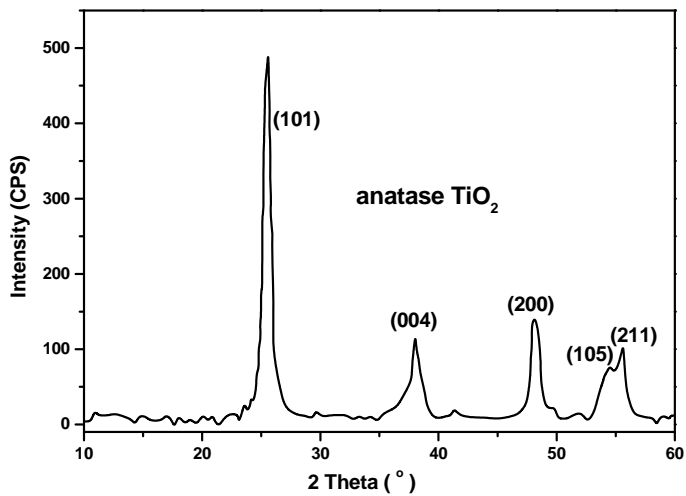

Figure 4. XRD pattern of the AP-BMI/Tit-5 nanocomposites.

*3.2. Morphology of the AP-BMI/TiO$_2$ nanocomposites.*

The morphology of the AP-BMI/TiO$_2$ nanocomposites has been studied by FE-SEM. Figure 5 gives the FE-SEM photographs of the fracture surface of the AP-BMI/TiO$_2$ nanocomposites. It can be seen that the AP-BMI resin matrix was a homogeneous material and no phase separation was observed (Figure 5, a), but the AP-BMI/TiO$_2$ nanocomposites were two-phase materials, in which the continuous phases were AP-BMI resin, and the dispersed phases were consisted of TiO$_2$ nanoparticles (Figure 5, b and c). The average diameters of primary particles of the dispersed phase in the nanocomposites were ca. 100nm, with particle aggregates of bigger sizes when the inorganic content increasing (Figure 5, b and c).

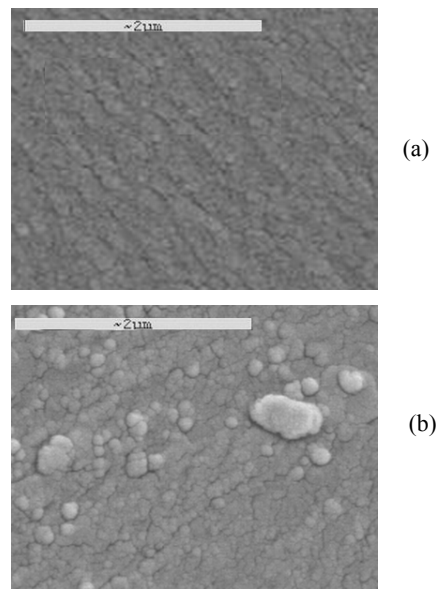

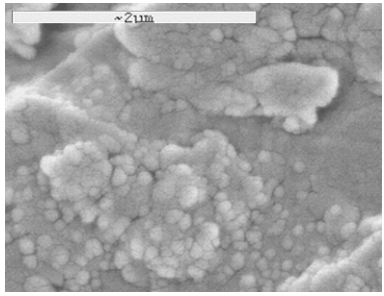

(c)

Figure 5. FE-SEM images for the AP-BMI resin/ TiO$_2$ nanocomposites (the scale bar is 2μm): (a) cured AP- BMI; (b) AP-BMI/Tit-2; and (c) AP-BMI/Tit-5.

*3.3. The thermal and mechanical properties of the the AP-BMI/TiO$_2$ nanocomposites.*
*3.3.1 $T_g$'s of the AP-BMI/TiO$_2$ nanocomposites.*

Glass transition temperatures ($T_g$'s) of the AP-BMI/TiO$_2$ nanocomposites are characterized by DMA, and the result is listed in Table 3. Figure 6 shows the $T_g$'s of the AP-BMI/TiO$_2$ nanocomposites change with inorganic content, based on the data from Table 3. It can be seen in Figure 6 that the incorporation of TiO$_2$ nanoparticles improved the glass transition temperature of the AP-BMI resin, due to the interaction of TiO$_2$ nanoparticles and AP-BMI molecular chain, where the TiO$_2$ behaves as physical cross-link points, or forms TiO$_2$/AP-BMI interpenetrating networks (through covalent bonds). Both factors will limit the molecular chain movement of AP-BMI resin, thus increase its glass transition temperature. At Zhao et al.' work [44], the introduction of TiO$_2$ also improved PEI-BMI resin's $T_g$, consistent with this paper, so it seems that the introduction of TiO$_2$ has obvious effect to increase $T_g$ of the resin in both works.

**Table 3.** Thermal resistantce of the AP-BMI resin/TiO$_2$ nanocomposites.

| Sample | Inorganic phase content (wt-%) | $T_g$ (°C) | Temp- for 5% wt- loss (°C) | Temp- for 10% wt- loss (°C) | $T_{max}$ (°C) | Wt- retention at 700°C (%) |
|---|---|---|---|---|---|---|
| AP-BMI | 0 | 272 | 415 | 452 | 655 | 42.5 |
| AP-BMI/Tit-2 | 2 | 278 | 436 | 471 | 650 | 47.4 |
| AP-BMI/Tit-5 | 5 | 302 | 436 | 512 | 635 | 51.6 |

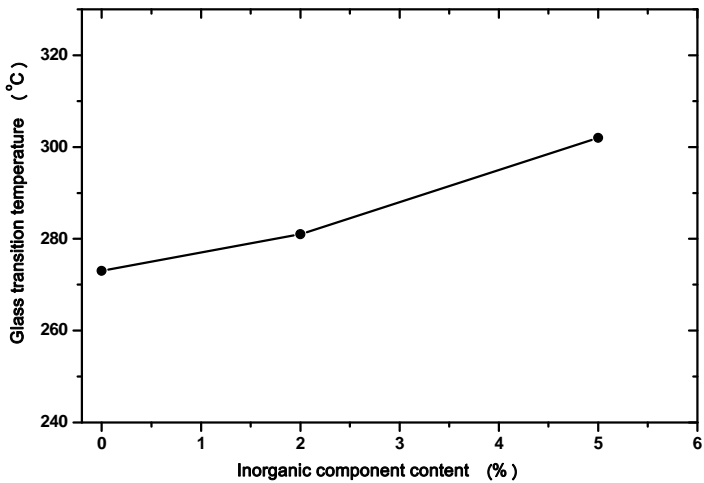

Figure 6. Glass transition temperatures of the AP-BMI resin/TiO$_2$ nanocomposites changes with inorganic component content.

*3.3.2 Thermal resistance of the AP-BMI TiO$_2$ nanocomposites.*

TGA diagrams of the AP-BMI/ TiO$_2$ nanocomposites with different inorganic phase contents are shown in Figure 7. It is seen from the figure that there were two stages in the weight-loss process of the nanocomposites. The elimination of uptake moisture accounted for the minor weight loss started at ~150°C. The major weight loss occurred at >450°C, which was due to the decomposition of the AP-BMI resin in the AP-BMI/TiO$_2$ nanocomposites. Table 3 summarizes the temperatures for 5% weight loss, 10% weight loss, and the temperatures at maximum decomposition rate ($T_{max}$) of the AP-BMI/ TiO$_2$ nanocomposites in the TGA diagrams.

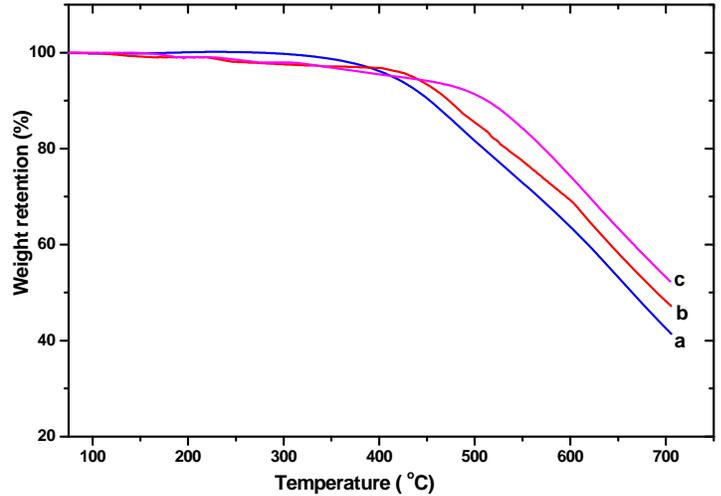

Figure 7. TGA diagrams of the AP-BMI resin/TiO$_2$ nanocomposites: (a) cured AP- BMI; (b) AP-BMI/Tit-2; and (c) AP-BMI/Tit-5.

It is seen that the introduction of the TiO$_2$ inorganic phase into the AP-BMI resin matrix improved the temperatures for 5% weight loss, 10% weight loss, and the weight retention at 700 °C of the resin, whereas the temperatures at maximum decomposition rate ($T_{max}$) have no big difference. The reasons accounted for the increase of the glass transition temperature of the AP-BMI/ TiO$_2$ nanocomposites also hold true for this situation.

*3.3.3 Dynamic moduli of the AP-BMI/TiO$_2$ nanocomposites.*

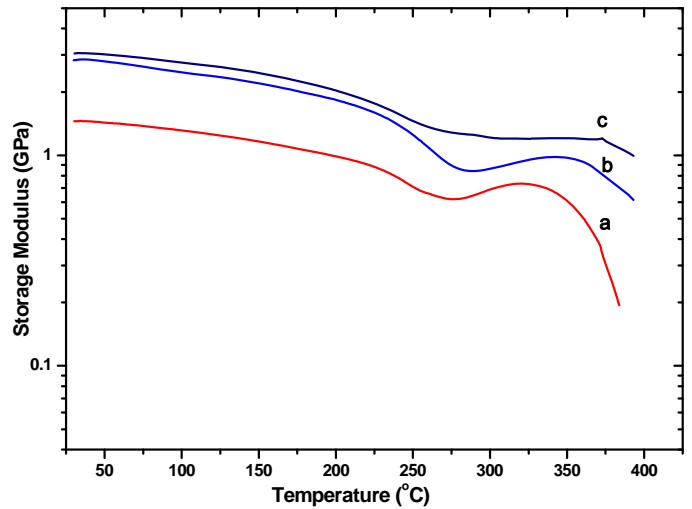

Figure 8. Change of storage modulus with temperature for the AP-BMI resin/TiO$_2$ nanocomposites: (a) cured AP- BMI; (b) AP-BMI-2/Tit; and (c) AP-BMI/Tit-5.

The change in storage modulus with temperature for the AP-BMI/TiO$_2$ nanocomposites is shown in Figure 8. It can be seen from the figure that the introduction of the titania phase into the AP-BMI resin improved the moduli of the nanocomposites. The main reason is that TiO$_2$ behaves as physical cross-link points in the nanocomposites, thus limiting the molecular movement of the AP-BMI resin.

## 4. Conclusions

The AP-BMI/TiO$_2$ nanocomposites have been prepared through the hydrolysis and condensation reactions of tetrabutyltitanate in the presence of AP-BMI prepolymers. The nanometer TiO$_2$ particles were formed in the BMI resin matrix and the average primary particle size of the dispersed phase in the nanocomposites was less than 100nm, but the particle aggregates with bigger size existed. The introduction of the nanosized TiO$_2$ inorganic phase into AP-BMI resin obviously improved $T_g$, the heat resistance of the BMI resin, and its dynamic modulus. The proposed mechanism accounted for the improved properties of the nanocomposites over AP-BMI resin is the interaction of TiO$_2$ nanoparticles and AP-BMI molecular chain, where the TiO$_2$ behaves as physical cross-link points, or forms TiO$_2$/AP-BMI interpenetrating networks.